# Experimental Assessment of Real-time PDCP-RLC V-RAN Split Transmission with 20 Gbit/s PAM4 Optical Access

A. El Ankouri[(1),(2)], L. Anet Neto[(1)], S. Barthomeuf[(1)], A. Sanhaji[(1)], B. Le Guyader[(1)], K. Grzybowski[(1)], S. Durel[(1)], P. Chanclou[(1)]

[(1)] Orange Labs, 2 Avenue Pierre Marzin - 22300 Lannion, France, anas.elankouri@orange.com
[(2)] IMT Atlantique, 655 Avenue du Technopôle, 29200 Plouzané

**Abstract:** *We experimentally demonstrate real-time, end-to-end transmission of 3GPP's option 2 functional split RAN interface with virtualized central units through up to 20km using a 20Gbit/s PAM4 link and 10GHz bandwidth optics.*

**Introduction**

Long Term Evolution was conceived following a clear trend to push the network intelligence towards its edges, with the whole radio protocol stack being processed in the Evolved Node B (eNBs) and the backhaul interface (S1) connecting the Evolved Packet Core (EPC) to the antenna sites (Fig. 1, top). Distributed Radio Access Networks (D-RAN) offered real advantages to the operators such as the ability to precisely target capacity increase needs[1].

Thanks to the rearrangement of the eNB functional blocks, we have witnessed later the emergence of centralized RAN (C-RAN) topologies, as opposed to D-RAN, with benefits such as reduced footprints at the antenna sites[2]. The main limitation of C-RAN is imposed, however, by the cost and availability of suitable low layer split fronthaul connectivity due to its stringent bit-rate and latency requirements[2].

It was clear that a new interface had to be conceived to accommodate the bandwidths expected for the 5G while allowing some degree of network centralization. Based on yet other distributions of the radio protocol stack, different functional splits were proposed and particularly fomented by the rise of software defined radio solutions. Indeed, those suit particularly well the highest layers of the radio stack, which are bounded to less strict latency constraints. This new virtual RAN (V-RAN) could enable a much faster optimization and evolution of the network thanks to easily (re)configurable and manageable instances on agnostic hardware.

Several possible splitting options have been defined by different standardization and industry groups. The 3GPP has defined a high-layer split interface, referred as V1 and F1 for the 4G and 5G respectively, between the Packet Data Convergence Protocol (PDCP) and Radio Link Control (RLC) blocks[3]. In Fig. 1 (bottom), a topology with a high layer split is shown where a central unit (CU) hosts virtualized layer 3 and part of layer 2 functions. The CU is connected to a distributed unit (DU) with a V1/F1 interface. The DU, with lower layer 2 and higher layer 1 blocks, is connected to the radio unit (RU), hosting the remainder of the PHY, through a low-layer split (not shown).

Previously, we have assessed a V1-ish interface in point-to-point and point-to-multipoint passive optical network (PON) topologies[4]. Here, we exploit a new solution based on an advanced modulation format for the optical access segment. Indeed, standardization activities on fixed optical access now focus on beyond XG(S)-PON[5] systems. New multi-level formats such as Pulse Amplitude Modulation (PAM) are good candidates to attain 20 Gbit/s with 10 GHz optics.

We experimentally demonstrate an end-to-end real-time transmission of a PDCP-RLC split interface through aggregation and access networks. The EPC and CU are virtualized and managed inside a virtualization environment in a host server. The aggregation segment between the virtual CU (vCU) and the switch in Fig. 1, is emulated by an Ethernet impairment engine that degrades the transmitted packets with variable latency, packet jitter and packet loss linked to bit error rate (BER). In the access segment, between the switch and the antenna site, we implement a physical layer transmission using real-time 20 Gbit/s PAM4 over 20 km of fibre.

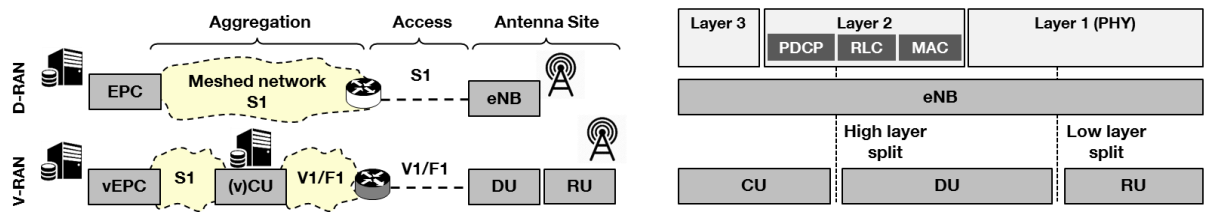

**Fig. 1:** D-RAN (top), C-RAN (center) and split V-RAN (bottom) topologies (left) and their splitting points (right).

**Experimental Setup**

Fig. 2 shows our experimental setup, which can be divided into 3 distinct parts:

Our radio plane runs on a server hosting the LTE mobile functions and is implemented on top of a single node CentOS Openstack virtualization environment. The EPC and radio protocol stack are aggregated in virtual machines, where each machine corresponds to a set of functions performed by an LTE node. For example, the EPC virtual machines, which offer the LTE core network services, contain the domain name system (DNS) server, the mobility management entity (MME), the serving and packet data network gateway (SPGW) and the home subscriber server (HSS). The EPC connects to the CU via the backhaul (S1) interface. The CU contains layer 3 and layer 2 up to the PDCP block of the LTE protocol stack. It generates a V1 interface, which goes out of the server over Ethernet and through the fixed aggregation and access networks before looping back to the same server. In our setup, we don't have a low-layer split interface and thus the DU and RU compose one single functional block. Also, the PHY layers of the RU and user equipment (UE) are abstracted. However, since our main objective is not to assess the mobile transmission through the air interface (Uu) but to evaluate the transmission of a high layer split through an optical transmission system, such abstraction can be made without loss of generality. The UE node is also implemented on a virtual machine and provisioned in the same server. It is important to notice that even though the various nodes are installed on the same server, they are logically separated and can only communicate via the existing mobile interfaces, also shown in Fig. 2

The second part of our experimental setup refers to the emulation of the aggregation network. This is done with a network impairment engine that can introduce latency, packet jitter and BER to the V1 interface. The V1 interface then goes to a 10 Gb Ethernet (10 GbE) switch that would also be connected to other cell sites in an actual deployed network. The traffic of those additional sites for both up and downlink (UL/DL) is created with an Ethernet traffic generator, allowing us to reach a symmetric throughput of 10.3125 Gbit/s. The overloading and V1 interface under evaluation are distinguished with different VLAN tags.

Finally, the fixed access plane is represented essentially by the real-time PAM4 encoder and decoder, the optical transceivers and a point-to-point transmission through 20 km of standard single mode fibre (SSMF). We use SFP+ (10G Small Form-factor Pluggable transceiver) modules and evaluation boards to provide connectivity adaptation between the 10 GbE switches and the inputs/outputs of our PAM4 bench. We focus only on the downlink. The uplink does not go through the PAM4 bench and is short circuited between the evaluation boards. We generate a de-correlated copy of the 10GbE stream containing the V1 payload and the overloading traffic and then we inject both streams as the most and least significant bits (MSB and LSB) inputs of our PAM4 encoder. The PAM4 signal is amplified with an electrical driver before modulating a 10 GHz Directly Modulated Laser (DML) emitting at 1311 nm. The optical signal goes through 20 km SSMF, representing the typical length of a Fixed Access Network segment. An attenuator adjusts the power at the input of an 8 GHz APD (Avalanche Photo Diode), with embedded transimpedance amplifier (TIA). The received electrical signal then attacks the PAM4 decoder, which separates the LSB and MSB flows of the PAM4 signal according to a previous report[6].

**Results and Discussions**

In order to assess the effects of the PAM4 modulation in our transmission, we take an optical back-to-back (OB2B) NRZ Ethernet transmission as reference and we compare it with the MSB flow of our PAM4 transmission after 20 km SSMF without forward error correction. To evaluate our access transmission and emulate different possible degradation phenomena coming from the aggregation network, we expressly degrade the V1 interface BER and we introduce a normally distributed latency variation in order to insert some packet jitter in our system.

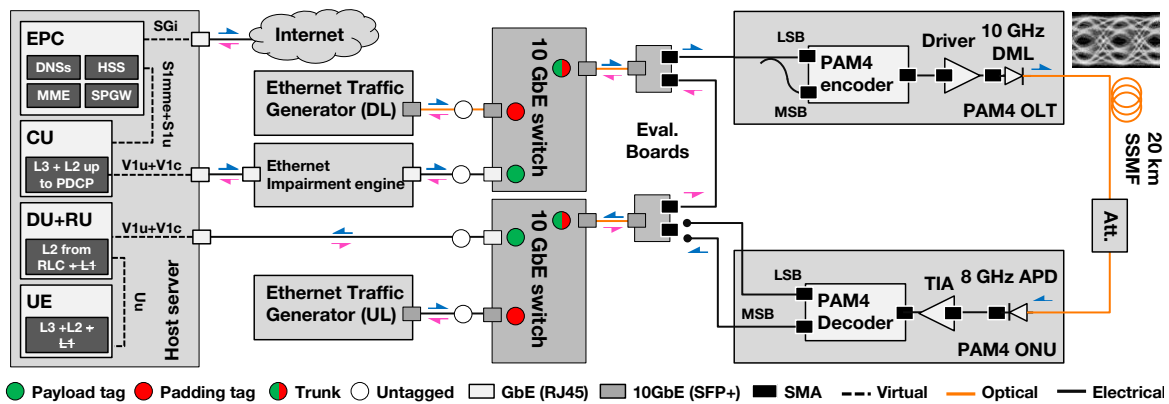

**Fig. 2:** Experimental setup.

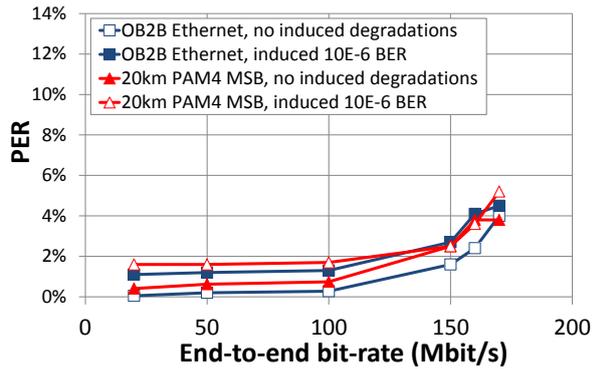

**Fig. 3:** PER variation with bitrate.

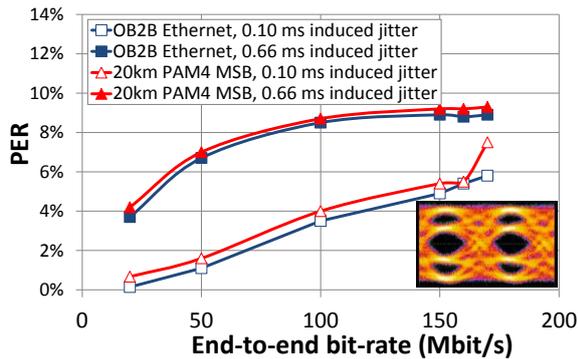

**Fig. 4:** PER vs bitrate for different induced jitters.

Fig. 3 shows the user datagram protocol (UDP) packet error rates (PER) variation between the EPC and the UE for different bit-rates and introduced BER values and for a packet size of 1200 bytes. We can see that the PER of our reference scenario is below 0.3% but increases for bit-rates beyond 150 Mb/s, which is due to limited resources in our virtual machines. We can also see that the PAM4 MSB optical access degrades the PER by ~0.8pp (percentage points) compared to the reference scenario. The introduction of a BER of $10^{-6}$ degrades the PER by about 0.5pp in both NRZ (Ethernet) and PAM4 transmissions compared their respective transmissions without degradation. In all cases, the PER remains below 5% for bit-rates up to 150 Mb/s, which correspond to the useful throughputs that can be transmitted in a 20 MHz, 2x2 multiple-input, multiple output (MIMO) LTE signal with 64QAM.

We have also measured the packet jitter between CU and DU with respect to the additional jitter introduced by our emulation engine (not show here for the sake of conciseness). We found out that the CU-DU jitter varies linearly with the induced jitter and that the additional packet jitter coming from the different equipment in our access transmission chain is ~120 µs. Also, the measured packet jitter values are roughly the same for the NRZ and PAM4 transmissions, meaning that the optical PHY signal jitter coming from the PAM modulation does not impact the packet jitter of the system.

Fig. 4 depicts the impact of the emulated packet jitter in the PER for different bit-rates, with an inset of the transmitted PAM4 eye-diagram. We fixed the mean introduced one-way latency to 2 ms and considered two values of packet jitter (latency standard deviation), namely 0.66 ms and 0.10 ms. The effect of the jitter is particularly noticeable and stronger for higher bit-rates. Whereas an introduced jitter of 0.1 ms imposes a linear PER degradation with respect to the bit-rate, a jitter of 0.66 ms imposes more abrupt signal degradation. For instance, we could measure ~4pp higher PER for an introduced packet jitter of 0.66 ms at 20 Mb/s and ~6pp for 50 Mb/s. Finally, the degradations introduced by the PAM4 modulation are relatively low compared to an ordinary Ethernet transmission. The measured packet jitter is 0.5pp and 0.3pp higher with the PAM4 for induced jitters of respectively 0.66 ms and 0.10 ms and bit-rates up to 150 Mb/s.

## Conclusions

In this work, we experimentally demonstrated the feasibility of transporting a high-layer mobile PDCP-RLC split interface over an Ethernet aggregation network and 20 km optical access network using real-time PAM4 modulation. We have also investigated the impacts of different impairments that could come from the aggregation network namely the BER and packet jitter.

## Acknowledgements

This work was supported by the European H2020-ICT-2016-2 project 5G-PHOS.